# Stress field modification near linear complexions increases the effective obstacle size and strengthening effect

Zhengyu Zhang [1], Daniel S. Gianola [2], Timothy J. Rupert [1,3,*]

[1] Hopkins Extreme Materials Institute, Johns Hopkins University, Baltimore, MD 21218, USA

[2] Materials Department, University of California, Santa Barbara, 93106, CA, USA

[3] Department of Materials Science and Engineering, Johns Hopkins University, Baltimore, MD 21218, USA

* Corresponding Author: tim.rupert@jhu.edu

**Abstract:**

Linear complexions are stable defect states that form along dislocations and recent experiments have demonstrated strengthening effects exceeding classical precipitation hardening predictions, motivating a detailed study of nanoscale strengthening mechanisms. Here, molecular dynamics simulations in Al-Cu and Ni-Al face-centered cubic alloys are used to demonstrate distinct plasticity mechanisms associated with linear complexions. Both nanoparticle array and platelet array complexions exhibit appreciable strengthening. In addition to direct interactions with the particles, stress field modification in nearby regions can restrict dislocation motion as well. Finally, the relative particle-dislocation orientation is found to have a large effect, with the strongest resistance observed when the dislocation stress field aligns with the original complexion nucleation condition. As a whole, these findings provide mechanistic insight into the strengthening observed experimentally and establish design principles for linear complexion-induced strengthening in structural alloys.

Complexions are thermodynamically-stable chemical and/or structural states confined to defects of all dimensionality, yet most attention has been focused on planar complexions at grain boundaries, interphase boundaries, or surfaces [1-12]. In particular, linear complexions (LCs) are an interesting case where segregation occurs near a line dislocation due to the local strain field until there is a local state transition [1, 3], representing a new pathway to modify the mechanical behavior of alloys. LCs were first discovered in body-centered cubic (BCC) alloys. Mn segregation to edge dislocations in Fe-Mn drove a local BCC to face-centered cubic (FCC) transformation confined to the dislocation core via spinodal fluctuations, immobilizing dislocations and producing discontinuous yielding that directly altered the mechanical response [5, 13]. For Fe-Ni alloys, atomistic simulations revealed that Ni segregation to edge dislocations drives the formation of core-shell $L1_0$ and B2 nanoscale precipitate arrays that serve as effective obstacles to dislocation motion [6, 14]. Unlike BCC alloys where only edge dislocations have hydrostatic stress components, all lattice dislocations in FCC materials dissociate into Shockley partial dislocations of mixed character, meaning every dislocation inherently has a non-zero hydrostatic stress field and can act as a potential LC nucleation site. Building on this, recent works have identified a range of possible LC types in FCC alloys [12, 15, 16], where Ni-Al and Al-Cu systems are of particular interest because they form nanoparticle arrays and platelet arrays, respectively. Notably, recent work by Howard et al. [16] shows extraordinary strengthening effects in Ni-Al that are significantly higher than expectations from classical theories [17-19], highlighting the unique and previously unrecognized contribution of LCs to mechanical performance. Importantly, these researchers observed small, chemically-ordered regions both at dislocations and elsewhere within the microstructure, demonstrating that linear complexion structures can persist even after the host dislocation has moved away. Dislocations would have to both break away from the linear complexions pinning them originally to begin to move as well as navigate through a network of retained linear complexions while continuing their glide. Collectively, these findings establish LCs as a promising strengthening mechanism that operates beyond classical precipitate and solid solution theories, motivating further investigation of LC formation and its effect on mechanical behavior.

The strengthening mechanisms associated with LCs are fundamentally distinct from typical precipitation hardening. In Al-Cu alloys, classical dislocation-precipitate interactions proceed through Orowan looping or particle shearing, with the dominant mechanism depending on precipitate size, orientation, and offset from the slip plane [20, 21]. In contrast, the local stress field associated with platelet array LCs forces dislocations into non-planar configurations through faceting and, in turn, local climb occurs near the platelets, requiring dislocations to climb down from the obstacle before glide can resume. This has been found to result in strain rate sensitivity values that far exceeds classical precipitate interactions [12, 22]. In Ni-Al alloys, classical $L1_2$ precipitate strengthening operates through particle shearing or Orowan bypassing, with a hybrid looping-shearing mechanism serving as the transition between the two regimes depending on precipitate size and volume fraction [23]. Nanoparticle array LCs in Ni-Al, however, resist dislocation motion through a fundamentally different pathway, where LC obstacles interact with dislocations through stress field modification even without crossing the slip plane [24]. Common to the two LC types described above is the fact that the LCs interact with dislocations through stress field modification. This stress field effect suggests that nearby gliding dislocations may interact with LCs at a distance without direct contact. In addition, the orientational character of LCs is inherited from the host Shockley partial dislocation, meaning the LC adopts a preferred orientation relative to the slip plane geometry. How this orientational nature influences dislocation-LC interactions is not understood.

In this study, we use molecular dynamics simulations to investigate LC strengthening mechanisms and address the gaps identified above. Both nanoparticle array and platelet array complexions are found to provide appreciable strengthening, with the effect spatially extending far beyond direct dislocation-particle contact. In addition, the relative orientation between the LC and the approaching dislocation is found to have a significant effect on strengthening. The largest strengthening occurs for dislocations of the same character as those which spawned the LC, yet a significant effect is still observed for dislocations of the opposite character. Together, these findings provide mechanistic insight into the experimentally observed strengthening associated with LCs.

Atomistic simulations were performed using the Large-scale Atomic/Molecular Massively Parallel Simulator (LAMMPS) package [25]. Structural characterization was carried out using common-neighbor analysis (CNA) [26], dislocation extraction analysis (DXA) [27] and polyhedral template matching (PTM) [27] implemented in OVITO [28]. When using DXA visualization, partial dislocations are shown as green lines; for PTM visualization, purple indicates unidentified structures ("Other"), blue represents pure FCC, green represents $L1_2$ A-sites and yellow-green denotes $L1_2$ B-sites. For both Al–Cu and Ni-Al system, LC samples were generated following a previously established hybrid Monte Carlo/molecular dynamics (MC/MD) protocol [29], with a time step of 1 fs for MD and one MC step performed for every 100 MD time steps. The equilibrated LCs were obtained using a Nose–Hoover thermostat and barostat at a constant temperature (250 K for Al-Cu and 300 K for Ni-Al) and zero external pressure with an initial pair of edge dislocations first relaxed to Shockley partials, with periodic boundary condition applied in all directions. Well-established embedded atom method potentials were used for Al-Cu [29] and Ni-Al [30] that are able to reproduce the phase diagrams of these alloy systems.

Equilibrium configurations for the LCs in the two alloys systems are presented in **Figure 1**. The X-axis of the samples is oriented along the [110] direction (Burgers vector of the original dislocation), the Y-axis oriented along the [111] direction (slip plane normal), and the Z-axis oriented along the [112] direction (line direction of the original dislocation). **Figures 1(a)** and **(b)** show 3D visualizations of LCs formed in the Ni-1.0 at.% Al and Al-0.3 at.% Cu systems, respectively, where Ni atoms are shown in red, Al atoms in blue, and Cu atoms in orange. **Figures 1(c) and (d)** shows PTM visualization for Ni-1.0 at.% Al and Al-0.3 at.% Cu, respectively. The Cu-rich LCs in Al-Cu are classified as "Other" in PTM due to their monolayer thickness, which prevents unambiguous structural template matching. The Ni-Al system forms a continuous nanoparticle array along the dislocation line, representing the extreme situation for this LC type where discrete particles have grown until they coalesce, while the Al-Cu system forms a discrete platelet morphology. The continuous array can be understood as a limiting case of the discrete obstacle configuration in which inter-particle spacing approaches zero [15].

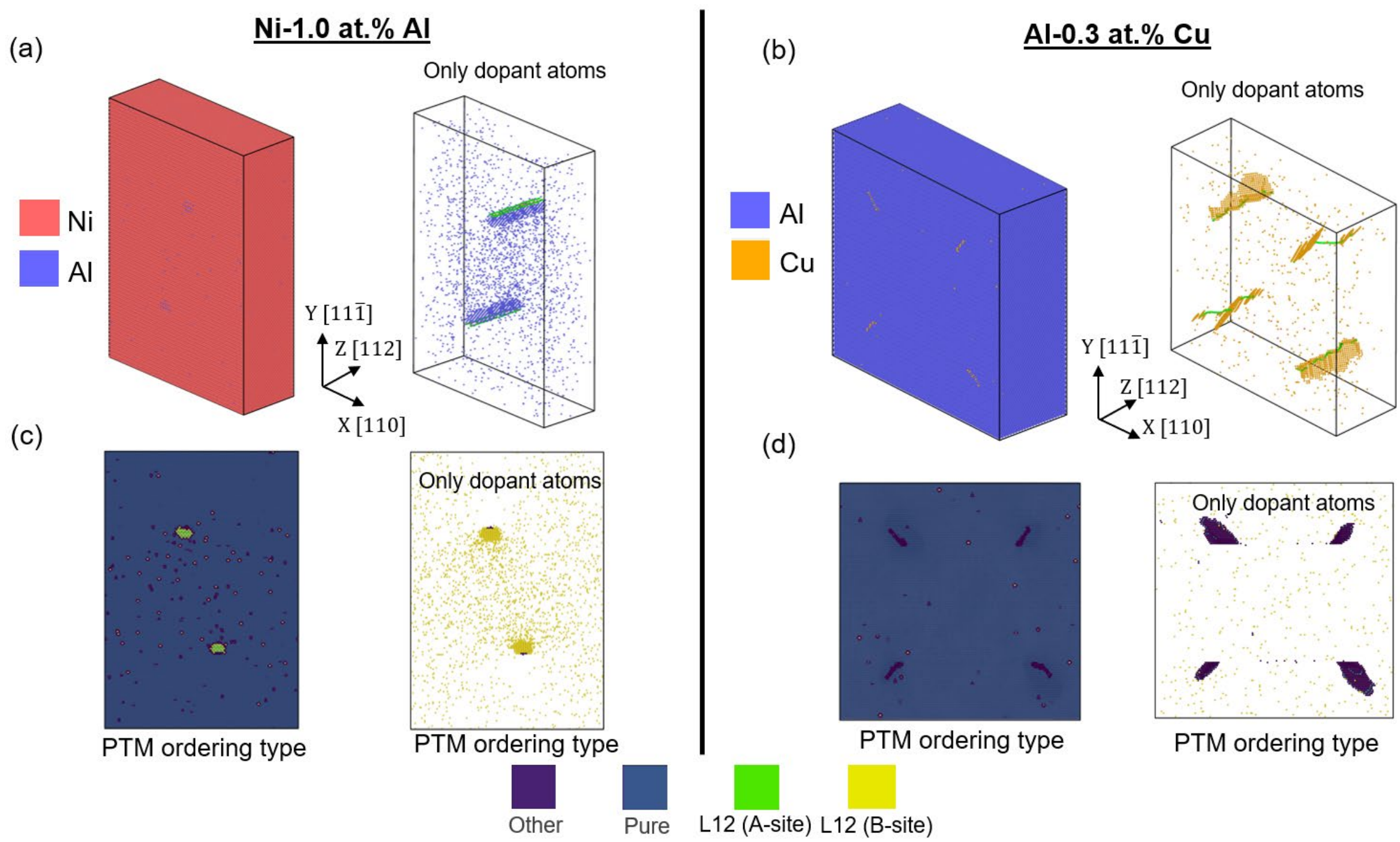

**Figure 1. Equilibrium configurations of linear complexions in Ni-Al and Al-Cu alloy systems. 3D visualization of LC structures formed by two pairs of Shockley partial dislocations after MC/MD equilibration: (a) Ni-1.0 at.% Al system with Ni atoms in red and Al atoms in blue; (b) Al-0.3 at.% Cu system with Al atoms in blue and Cu atoms in orange. The panels on the left show the full simulation cell, while those on the right display only dopant atoms for clarity and the Shockley partials appear as green lines. PTM analysis of the close-packed plane view for (c) Ni-Al and (d) Al-Cu. Panels on the left show the full view, while panels on the right show dopant atoms only and the Shockley partials appear as green lines.**

To create simulation cells for shear deformation simulations, one LC (particles only) was isolated for each alloy system. A new edge dislocation was introduced on a (111) close-packed by cutting a half plane normal to the [110] Burgers vector direction, with the exact plane location with respect to the original glide plane varied to probe different configurations. Next, a Nose–Hoover thermostat at 250K (Al-Cu) or 300K (Ni-Al) and a barostat at zero external pressure were applied for another 100 ps. During shear deformation, the boundary condition in the Y-direction was changed to non-periodic, with fixed boundary layers applied at the top and bottom surfaces to avoid elastic interaction effects that would have occurred if a dipole configuration had been used. Accurate representation of mechanical behavior in shear simulations requires proper selection of interatomic potentials. The Ni-Al potential used in LCs growth was also applied in the shear test, as it accurately reproduces important material properties such as elastic constants, defect energies, major stable phases and used in previous study [15]. While the Al-Cu interatomic potential mentioned above was developed to reproduce important features of the bulk phase diagram and therefore correctly predicts second phase formation in Al-Cu [12, 22], it significantly underestimates the stacking fault energy of Al. As such, a second angular dependent interatomic potential from Apostol and Mishin [31] that accurately predicts stacking fault energy and other important mechanical properties was used for deformation simulations. Constant shear stress was applied to the Y-axis faces in the X-direction. Then, the critical stress for de-pinning was determined by changing the applied stress until the dislocation remained stationary for at least 50 ps within an NVT simulation held at 250K (Al-Cu) or 300K (Ni-Al). Both elemental potentials accurately reproduce lattice parameters, elastic constants, stacking fault energies, and phase stabilities obtained from experimental studies and first-principles calculations. Further details of potential validation are provided in Supplementary Note 1.

The LC formation geometry, dislocation orientation definitions, and a schematic of the shear test for both alloy systems are illustrated in **Figure 2**. In the Ni-Al system (**Figure 2(a)**), Al segregates to the tensile side of the dislocations, forming $L1_2$ particles on the tensile side. In contrast, the Al-Cu system (**Figure 2(b)**) experiences Cu segregation to the compressive side. The formation of these particles would not be expected in the absence of dislocations since both compositions are intentionally selected to reside in the single phase-field region of the bulk phase diagram. In plastically deformed FCC metals, edge dislocation dipoles consisting of pairs of dislocations with opposite signs on adjacent slip planes are a natural and ubiquitous product of deformation [32, 33]. Because of this, both orientations of the newly introduced (glissile) dislocation must be considered, as shown in the middle panel of **Figure 2**. Among these two orientations, one is favored to lower the total strain energy ("Favored") because it is the orientation that created the LC while the other has the opposite character ("Non-favored"), with this effect possible creating an anisotropic interaction between the LC and any incoming dislocation. The right panels of **Figure 2** show schematic representations of the close-packed XZ plane, with the dislocation position with respect to the LC shown. For convenience, we define $d = 0$ as the slip plane just above or below the LC, giving two reference planes for each system. The distance $d$ is then measured in the [111] direction from each reference plane. Negative values indicate positions away from the location where the LC was originally created, while positive positions denote distances from the opposite side of the LC. Shear tests were conducted at various distances for both orientations to completely map any strengthening effect range of LCs.

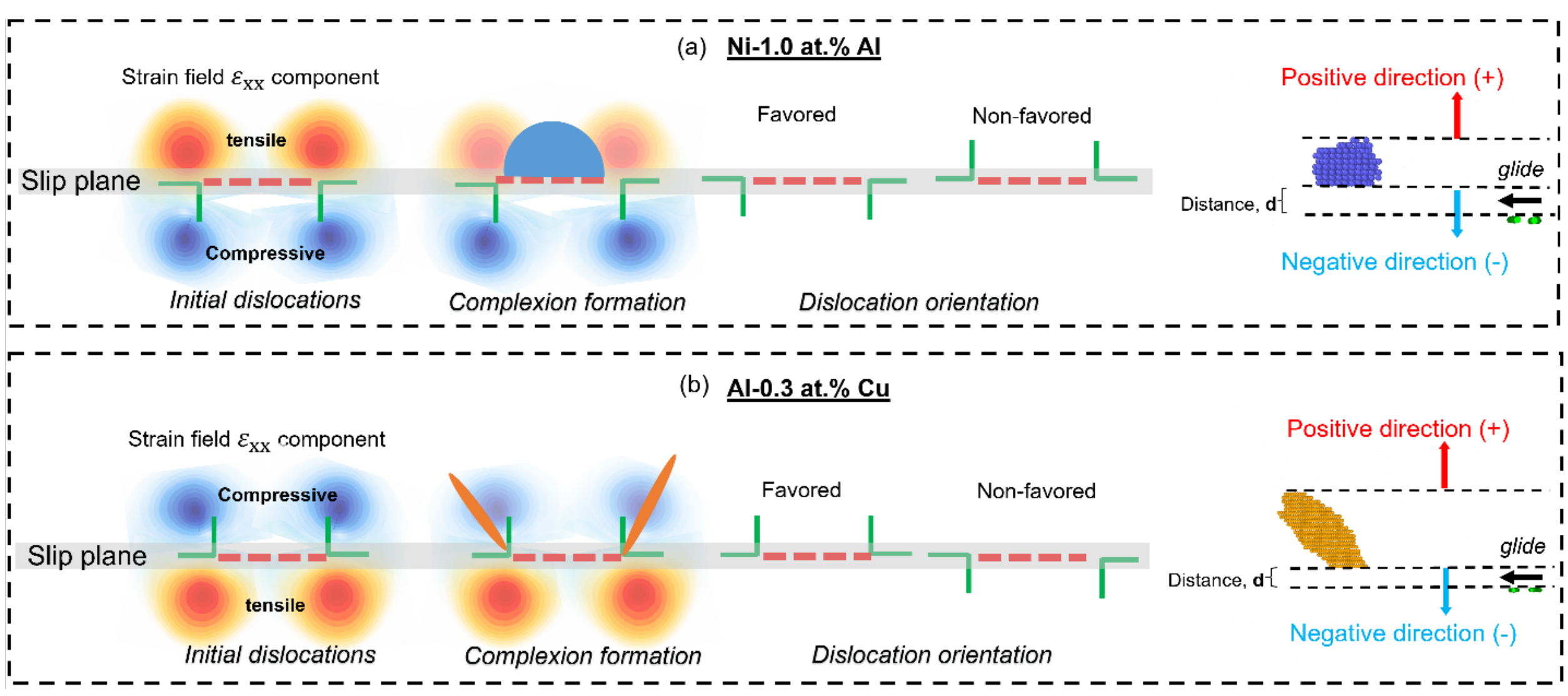

**Figure 2. Schematic illustration of strain field evolution during linear complexion formation and the resulting orientation-dependent dislocation interactions. (a) In Ni-Al system, Al segregates to the tensile side and forms $L1_2$ nano particles. (b) In Al-Cu, Cu segregates to the compressive side of the Shockley partial dislocations and forms platelet complexions. The left panels show the $\varepsilon_{xx}$ strain field component of the initial dislocation with LC formation in the compressive (blue) and tensile (red) regions denoted. The middle panels depict the Favored Non-favored configuration for each LC. The right panels show the testing geometry, where distance, *d*, is measured along the [111] direction from the slip plane.**

The applied shear stress required to depin the dislocation from the LC, hereafter referred to as the critical shear stress, as a function of dislocation distance is presented in **Figure 3**, with the blue dashed line indicating the solid solution strengthening baseline (obtained for the same simulation cell without distinct LCs). Both alloys show an asymmetric critical shear stress distribution across the reference planes, with the stronger interactions occurring near the location of the LC formation (i.e., negative values). Importantly though, there is substantial strengthening on both sides that extends significantly beyond direct contact with the LC. In Ni-Al (**Figure 3a**), the nanoparticle LC produces substantially higher peak stresses, where the Favored orientation reaches a maximum of 1.2 GPa (at $d = 0$) and the Non-favored orientation 0.8 GPa (at $d = -0.3$ nm). Critical stress remaining above 0.6 GPa over ~1.2 nm on the complexion formation side (right) and ~0.6 nm on the opposite side (left), for at least one dislocation orientation. In Al-Cu (**Figure 3b**), the peak critical stress reaches a maximum value of 0.32 GPa for the Favored orientation and 0.18 GPa for the Non-favored orientation at $d = 0$, with the strengthening effect again extending approximately a nanometer on each side. Coherency stress field interactions have been predicted for classical precipitates [34], and the strengthening effect from the LCs can be viewed as a similar form of stress field modification. Coherency stresses and their resultant strengthening effect should fall off rapidly as distance from the precipitate is increased, which is consistent with the data shown here for LCs in Figure 3. There are important differences for the two alloys. In Ni-Al, the Favored orientation is both stronger and has a more spatially-sustained effect, maintaining the elevated critical stress over greater distances. In contrast, the Favored orientation produces stronger pinning immediately adjacent to the platelet array LC in Al-Cu, yet the Non-favored orientation is stronger once the dislocation is only a few angstroms away. The LCs simulated here were chosen to be similar in size to those observed by Howard et al. [16]. It is important to note that LCs are only rigorously stable when the dislocation's stress field is nearby, so some level of dissolution is possible after the original unpinning event. This possibility is discussed in Supplementary Note 2, where LCs maintain a considerable strengthening effect above the solid solution baseline in both alloy systems even if there is significant shrinkage of the original particle. Hence, even if some amount of dissolution occurred from the stable LC state, one would still expect a significant strengthening effect.

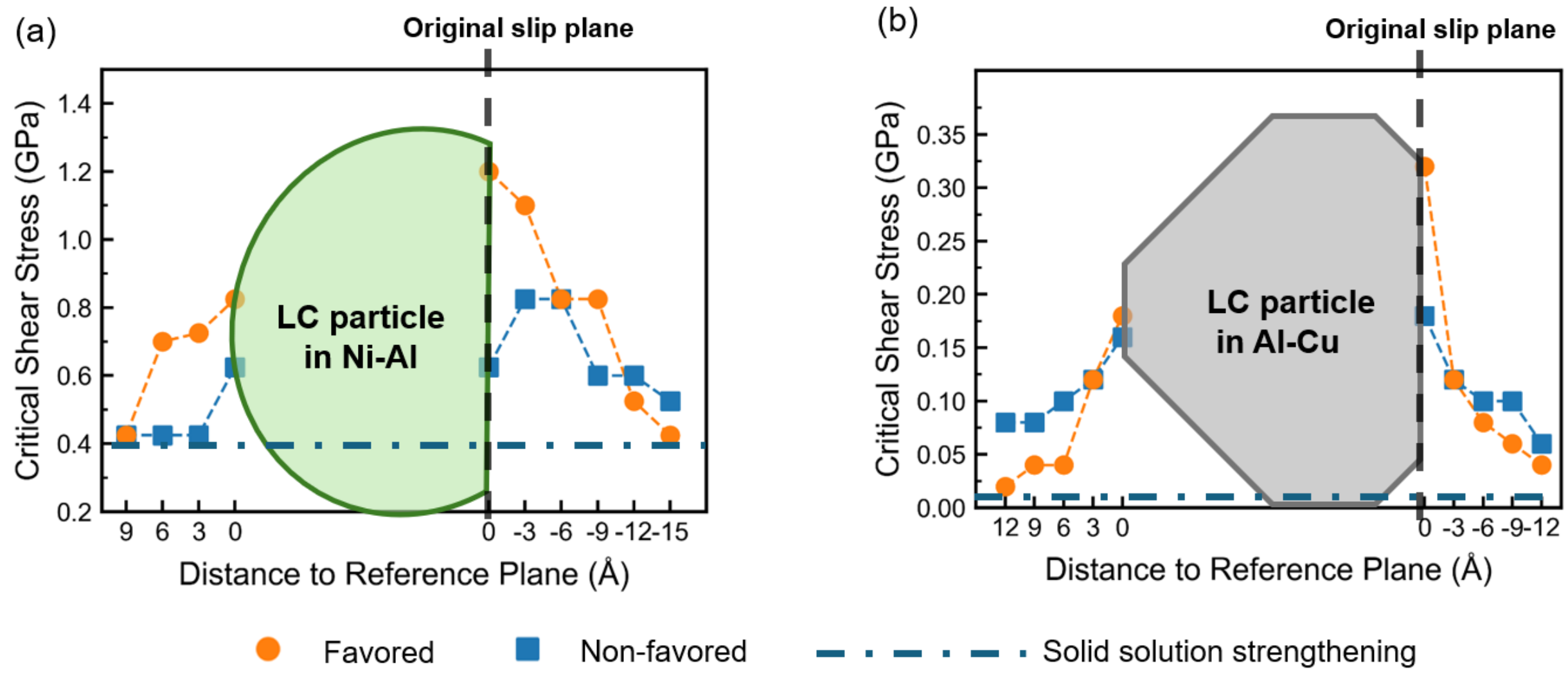

**Figure 3. Critical shear stress as a function of distance from the original slip plane for (a) Ni-Al with nanoparticle array LCs and (b) Al-Cu with platelet array LCs. Orange circles represent the Favored dislocation orientation and blue squares represent the Non-favored orientation. The green and gray shaded regions indicate the LCs for the Ni-Al and Al-Cu alloys, respectively. The blue dash-dot line represents the**

**solid solution strengthening threshold, below which dislocation motion occurs in simulations without LC. The vertical dashed line marks the original slip plane of the parent partial dislocations.**

**Figure 4** shows the atomic configurations at maximum critical stress for a distance of -1.2 nm (slightly below where the complexion formed). In both the Al-Cu and Ni-Al alloys, Favored dislocations are pinned directly below the LCs (**Figures 4(a) and (b)**), while Non-favored dislocations are repelled and held at remote positions on the slip plane (**Figures 4(c) and(d)**). The latter demonstrates that LC strengthening is similar to solid solution strengthening in that it does not matter if the dislocation is attracted to or repelled by the obstacle, as the collective pathway for motion becomes harder to traverse. Importantly, both pinning effects show that long-range elastic interactions can immobilize dislocations without direct contact, highlighting the stress field modification effect. As demonstrated by the Non-favored case, even dislocations of different character can be pinned by nearby complexions without direct contact, providing a mechanistic explanation for the large strengthening effects observed experimentally [16].

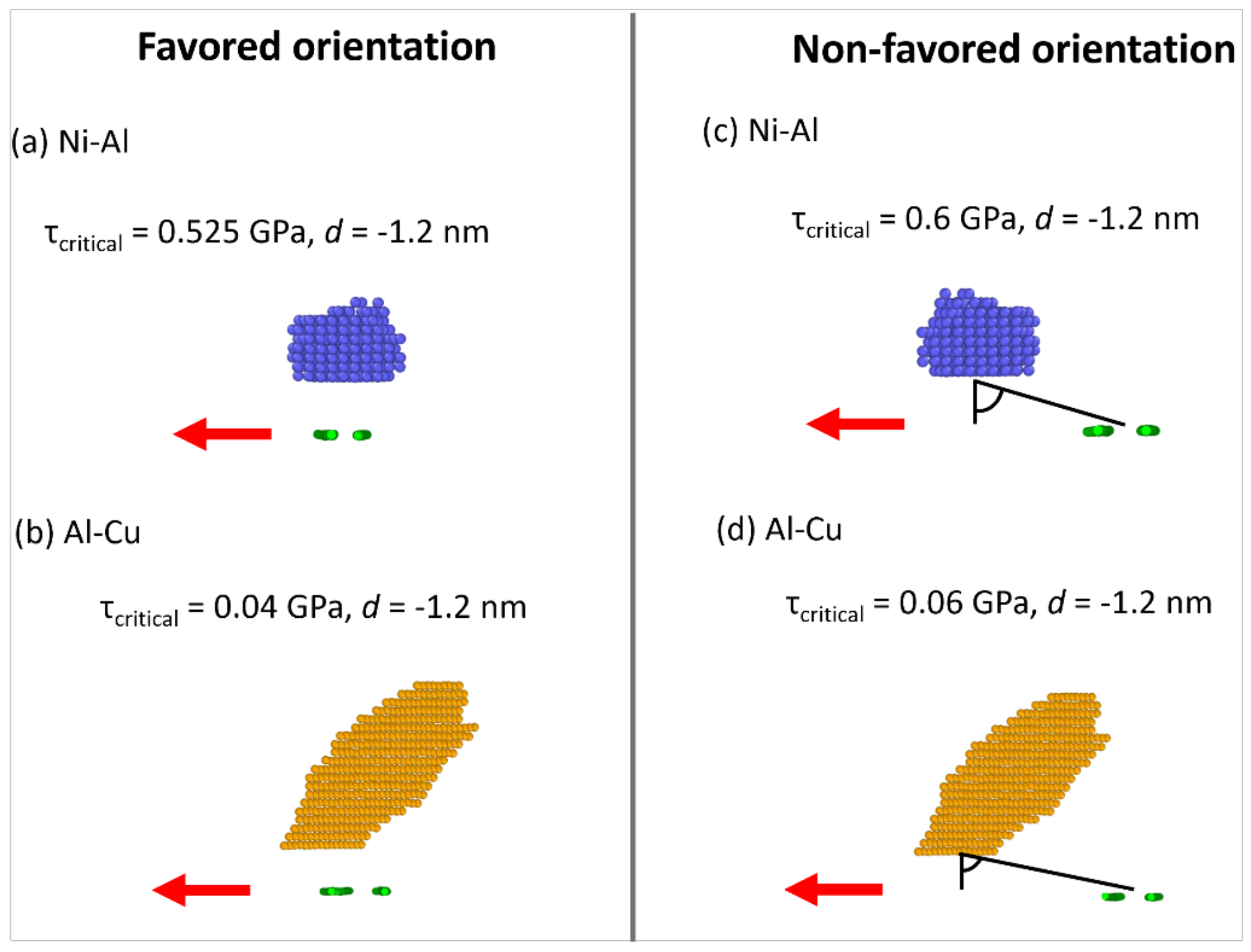

**Figure 4. Atomic configurations at maximum critical stress for both dislocation orientations for *d* = -1.2 nm. Favored orientations for (a) Ni-Al and (b) Al-Cu systems show dislocations pinned directly underneath the LCs. Non-favored orientations for (c) Ni-Al and (d) Al-Cu show dislocations held at remote positions away from the LCs due to strain field incompatibility. Solvent atoms are removed and dislocations are identified by DXA.**

The pinning configurations revealed in **Figure 4** indicate that dislocations can interact with the LCs structure at remote distances, motivating a quantitative analysis of how this extended interaction modifies classical strengthening predictions. For the Ni-Al system, anti-phase boundary (APB) strengthening [17, 18] is relevant here as a comparison, as the LC region is associated with local $L1_2$ ordered domains that dislocations would have to shear if they came into direct contact with the particles. The APB strengthening model captures the thermodynamic cost of a dislocation cutting through an ordered boundary in terms of energy-based measure that is independent of the specific bypass mechanism. Since the strengthening resistance is derived from the spatial gradient of interaction energy, an extended dislocation-LC interaction zone is equivalent to increasing the effective obstacle size in classical APB strengthening models. APB strengthening analysis can be used as a first approximation for particle strengthening in this system:

$$\Delta\sigma_{OS} = M \times 0.81 \times \frac{\gamma_{APB}}{2b} \times \left(\frac{3\pi f}{8}\right)^{\frac{1}{2}} \tag{1}$$

where M is the Taylor factor, $\gamma_{APB}$ is the APB energy, $b$ is the Burgers vector magnitude, and $f$ is the volume fraction. **Figure 5** schematically illustrates how a collection of classical particles (shown in blue) would hinder dislocation glide along random planes (grey dashed lines), with the obstacle fraction (with the strengthening effect being proportional) for each plane projected onto the plot on the right. For this exercise, particle positions are first generated using an MC procedure to give a random distribution and the effective diameters were chosen as 2.7 ± 0.59 nm to match experimental observations in Ni-Al from Howard et al. [16].

However, this classical treatment assumes dislocation-particle interaction only occurs at direct contact, and we therefore incorporate the extended stress modification effect observed in **Figure 3** to quantify how the effective particle size and resulting strengthening would be modified. Using the Ni-Al system as an example, a substantial strengthening effect (subjectively defined as a 50% increase over the solid solution lower bound in **Figure 3**) is observed for 1.2 nm below the LC and 0.6 nm above the nanoparticles. If this total 1.8 nm is added to the effective obstacle diameter, the obstacle size for LCs would be 4.5 ± 0.59 nm or a diameter increase of 67% from the direct contact case. Coherent precipitates can also have coherency strain effects, yet these particles are often much larger than the LCs studied here. For example, classical coherent $L1_2$ precipitates typically exhibit optimal strengthening at sizes of ~20–60 nm [35-37]. Even with refractory element additions such as Ta chosen to restrict diffusion, particle sizes remain ~7 nm under optimized aging conditions [38]. In contrast, the LCs have widths along the dislocation line of only 3 nm in the Ni-Al alloy and 4 nm in the Al-Cu alloy. The size of these LCs is inherently limited by the size of the high stress region near the dislocation, keeping them very small and ensuring that a strong stress modification effect occurs.

The linear complexion case (orange lines in **Figure 5**) exhibits consistently higher obstacle fraction than the classical particle case, demonstrating that extended strain-field interactions increase the effective barrier to dislocation motion beyond what geometric contact alone would predict. This increased size would lead to a 116% increase in strengthening effect increase following the cubic root dependence of volume fraction on particle diameter, *D*:

$$\Delta\sigma_{OS} \propto f^{\frac{1}{2}} \propto D^{\frac{3}{2}} \quad (2)$$

The extended effective obstacle size therefore results in a much larger strengthening effect. In the analysis of Howard et al. [16], APB-based particle shearing theory predicted a strength of 253–338 MPa while the experimentally measured strength was 814 MPa. The LC-induced strength increase of roughly 140-220% beyond what conventional mechanisms can account for is similar in magnitude to the predictions made here (~116% increase in strength).

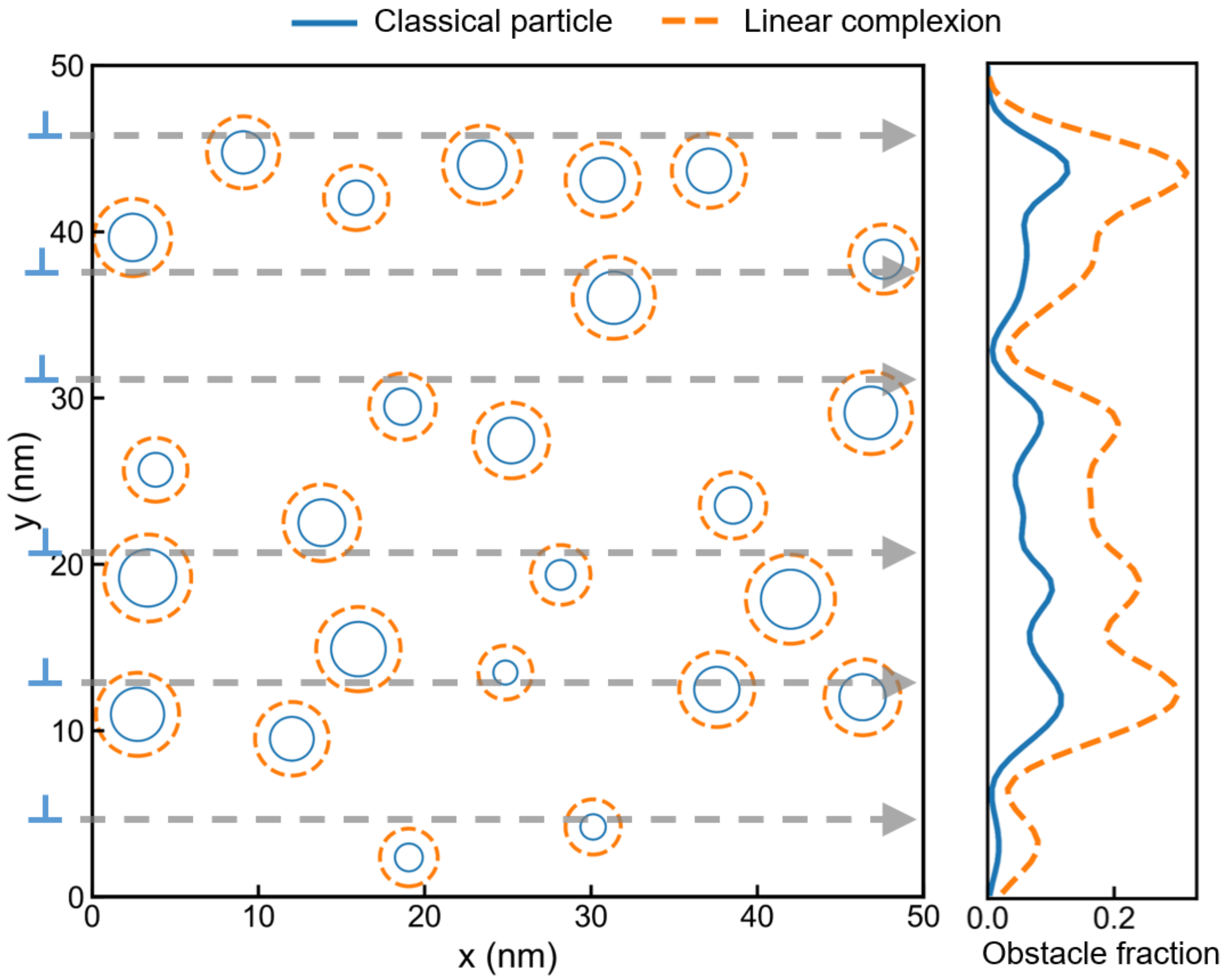

**Figure 5. Schematic illustration of dislocation strengthening from classical particles and LCs. Left panel: Particle positions generated by a Monte Carlo procedure within a 50 nm × 50 nm cell, with diameters of 2.7 nm for classical particles (blue circles) and 4.5 nm for LCs (dashed orange circles), to account for the stress field modification effect. Horizontal dashed lines with edge dislocation symbols indicate potential glide paths, with grey arrows denoting glide direction. Right panel: Obstacle fraction along each glide plane, defined as the total length of obstacles a dislocation need to overcome divided by the total glide path length at each Y-position.**

In summary, this work investigated dislocation-LC interactions in both nanoparticle array (Ni-Al) and platelet array (Al-Cu) systems using atomistic simulations, revealing strengthening mechanisms that were much more effective than classical precipitation hardening. First, both LC types produce critical shear stress distributions that extend significantly beyond the geometric boundaries of the particles, with the strongest effect occurring on the side and with the dislocation orientation where the complexion was originally created. Second, the dislocation-LC interactions are strongly orientation-dependent, with Favored dislocation orientations attracted to the LC core and Non-favored orientations repelled. However, both result in strengthening. Finally, incorporation of the stress field modification effect into a simple classical strengthening model predicts a 116% strengthening increase relative to classical direct-contact predictions, demonstrating the effectiveness of LCs for strengthening.

**Acknowledgements**

This research was sponsored by the Army Research Office under Grant Number W911NF-21–1–0288. The views and conclusions contained in this document are those of the authors and should not be interpreted as representing the official policies, either expressed or implied, of the Army Research Office or the U.S. Government. The U.S. Government is authorized to reproduce and distribute reprints for Government purposes notwithstanding any copyright notation herein.

## References

[1] M. Kuzmina, M. Herbig, D. Ponge, S. Sandlöbes, D. Raabe, Linear complexions: Confined chemical and structural states at dislocations, Science 349(6252) (2015) 1080-1083.
[2] S.J. Dillon, M. Tang, W.C. Carter, M.P. Harmer, Complexion: a new concept for kinetic engineering in materials science, Acta Materialia 55(18) (2007) 6208-6218.
[3] P.R. Cantwell, M. Tang, S.J. Dillon, J. Luo, G.S. Rohrer, M.P. Harmer, Grain boundary complexions, Acta Materialia 62 (2014) 1-48.
[4] J. Luo, Grain boundary complexions: the interplay of premelting, prewetting, and multilayer adsorption, Applied Physics Letters 95(7) (2009).
[5] A. Kwiatkowski da Silva, D. Ponge, Z. Peng, G. Inden, Y. Lu, A. Breen, B. Gault, D. Raabe, Phase nucleation through confined spinodal fluctuations at crystal defects evidenced in Fe-Mn alloys, Nature communications 9(1) (2018) 1137.
[6] V. Turlo, T.J. Rupert, Linear complexions: Metastable phase formation and coexistence at dislocations, Physical Review Letters 122(12) (2019) 126102.
[7] V. Turlo, T.J. Rupert, Prediction of a wide variety of linear complexions in face centered cubic alloys, Acta Materialia 185 (2020) 129-141.
[8] J. Luo, X. Shi, Grain boundary disordering in binary alloys, Applied Physics Letters 92(10) (2008).
[9] D. Raabe, M. Herbig, S. Sandlöbes, Y. Li, D. Tytko, M. Kuzmina, D. Ponge, P.-P. Choi, Grain boundary segregation engineering in metallic alloys: A pathway to the design of interfaces, Current Opinion in Solid State and Materials Science 18(4) (2014) 253-261.
[10] C.H. Liebscher, A. Stoffers, M. Alam, L. Lymperakis, O. Cojocaru-Mirédin, B. Gault, J. Neugebauer, G. Dehm, C. Scheu, D. Raabe, Strain-induced asymmetric line segregation at faceted Si grain boundaries, Physical review letters 121(1) (2018) 015702.
[11] T. Frolov, M. Asta, Y. Mishin, Segregation-induced phase transformations in grain boundaries, Physical Review B 92(2) (2015) 020103.
[12] P. Garg, D.S. Gianola, T.J. Rupert, Enhanced strain rate sensitivity due to platelet linear complexions in Al-Cu, Scripta Materialia 271 (2026) 117002.
[13] A.K. Da Silva, G. Leyson, M. Kuzmina, D. Ponge, M. Herbig, S. Sandlöbes, B. Gault, J. Neugebauer, D. Raabe, Confined chemical and structural states at dislocations in Fe-9wt% Mn steels: A correlative TEM-atom probe study combined with multiscale modelling, Acta Materialia 124 (2017) 305-315.
[14] V. Turlo, T.J. Rupert, Dislocation-assisted linear complexion formation driven by segregation, Scripta Materialia 154 (2018) 25-29.
[15] D. Singh, D.S. Gianola, T.J. Rupert, Dislocation breakaway from nanoparticle array linear complexions: Plasticity mechanisms and strength scaling laws, Materialia 32 (2023) 101929.
[16] H. Howard, W. Cunningham, A. Genc, B. Rhodes, B. Merle, T. Rupert, D. Gianola, Chemically ordered dislocation defect phases as a new strengthening pathway in Ni–Al alloys, Acta Materialia 289 (2025) 120887.
[17] Q. Wang, Z. Li, S. Pang, X. Li, C. Dong, P.K. Liaw, Coherent precipitation and strengthening in compositionally complex alloys: a review, Entropy 20(11) (2018) 878.
[18] A.J. Ardell, Precipitation hardening, Metallurgical Transactions A 16(12) (1985) 2131-2165.
[19] W. Cai, W.D. Nix, Imperfections in crystalline solids, Cambridge University Press2016.
[20] C. Singh, D. Warner, Mechanisms of Guinier–Preston zone hardening in the athermal limit, Acta Materialia 58(17) (2010) 5797-5805.

[21] C. Singh, A. Mateos, D. Warner, Atomistic simulations of dislocation–precipitate interactions emphasize importance of cross-slip, Scripta Materialia 64(5) (2011) 398-401.
[22] P. Garg, D.S. Gianola, T.J. Rupert, Strengthening from dislocation restructuring and local climb at platelet linear complexions in Al-Cu alloys, Journal of Materials Science: Materials Theory 8(1) (2024) 9.
[23] S. Chatterjee, Y. Li, G. Po, A discrete dislocation dynamics study of precipitate bypass mechanisms in nickel-based superalloys, International journal of plasticity 145 (2021) 103062.
[24] D. Singh, V. Turlo, D.S. Gianola, T.J. Rupert, Linear complexions directly modify dislocation motion in face-centered cubic alloys, Materials Science and Engineering: A 870 (2023) 144875.
[25] A.P. Thompson, H.M. Aktulga, R. Berger, D.S. Bolintineanu, W.M. Brown, P.S. Crozier, P.J. In't Veld, A. Kohlmeyer, S.G. Moore, T.D. Nguyen, LAMMPS-a flexible simulation tool for particle-based materials modeling at the atomic, meso, and continuum scales, Computer physics communications 271 (2022) 108171.
[26] D. Faken, H. Jónsson, Systematic analysis of local atomic structure combined with 3D computer graphics, Computational Materials Science 2(2) (1994) 279-286.
[27] A. Stukowski, V.V. Bulatov, A. Arsenlis, Automated identification and indexing of dislocations in crystal interfaces, Modelling and Simulation in Materials Science and Engineering 20(8) (2012) 085007.
[28] A. Stukowski, Visualization and analysis of atomistic simulation data with OVITO–the Open Visualization Tool, Modelling and simulation in materials science and engineering 18(1) (2009) 015012.
[29] Y. Cheng, E. Ma, H. Sheng, Atomic level structure in multicomponent bulk metallic glass, Physical review letters 102(24) (2009) 245501.
[30] Y. Mishin, M. Mehl, D. Papaconstantopoulos, Embedded-atom potential for B 2− NiAl, Physical review B 65(22) (2002) 224114.
[31] F. Apostol, Y. Mishin, Interatomic potential for the Al-Cu system, Physical Review B—Condensed Matter and Materials Physics 83(5) (2011) 054116.
[32] P. Neumann, Low energy dislocation configurations: a possible key to the understanding of fatigue, Materials Science and Engineering 81 (1986) 465-475.
[33] A. Aslanides, V. Pontikis, Numerical study of the athermal annihilation of edge-dislocation dipoles, Philosophical Magazine A 80(10) (2000) 2337-2353.
[34] V. Gerold, H. Haberkorn, On the critical resolved shear stress of solid solutions containing coherent precipitates, physica status solidi (b) 16(2) (1966) 675-684.
[35] J. He, H. Wang, H. Huang, X. Xu, M. Chen, Y. Wu, X. Liu, T. Nieh, K. An, Z. Lu, A precipitation-hardened high-entropy alloy with outstanding tensile properties, Acta Materialia 102 (2016) 187-196.
[36] Y. Zhao, H. Chen, Z. Lu, T. Nieh, Thermal stability and coarsening of coherent particles in a precipitation-hardened (NiCoFeCr) 94Ti2Al4 high-entropy alloy, Acta Materialia 147 (2018) 184-194.
[37] G. Qin, Z. Zhang, W. Wang, X. Wu, F. Yuan, Tailoring size and fraction of coherent L12 nanoprecipitates to achieve strong hardening in medium entropy alloys with heterogeneous grain structures, Journal of Materials Research and Technology 32 (2024) 2963-2971.
[38] D. Zhang, J. Kuang, H. Xue, J. Zhang, G. Liu, J. Sun, A strong and ductile NiCoCr-based medium-entropy alloy strengthened by coherent nanoparticles with superb thermal-stability, Journal of Materials Science & Technology 132 (2023) 201-212.

## Supplementary Materials

### Supplementary Note 1: Interatomic potentials used in shear simulations

For the shear simulations of Ni-Al, the interatomic potential used in this work is the embedded-atom method (EAM) potential for Ni-Al developed by Mishin et al. [1], which was constructed by fitting to both experimental properties and ab initio calculations using the linearized augmented-plane-wave method (LAPW) across a wide range of structures and configurations. The potential accurately reproduces the equilibrium lattice parameter of B2-NiAl (2.86 Å vs. experimental 2.88 Å [2]) and $L1_2$-$Ni_3Al$ (3.525 Å vs. experimental 3.57 Å [3]). For phase stability, the B2 and $L1_2$ structures are correctly predicted to be the lowest-energy structures of NiAl and $Ni_3Al$ respectively, with good agreement between the EAM and LAPW calculations verified across multiple alternative structures including B1, $L1_0$, $L1_1$, B32, $DO_{22}$, and D03. Regarding $L1_2$ faulting and slip system selection, the potential gives a fully relaxed anti-phase boundary (APB) energy of 0.55 J/m² on the (110) plane, consistent with the experimental lower bound of 0.50 J/m² [4]. This correctly reproduces the preference for [001] slip over [111] slip that is observed experimentally in B2-NiAl. These properties collectively confirm that the potential is well-suited for studying the structure, stability, and mechanical response of the Ni-Al LCs investigated in this work.

The ADP potential for the Al-Cu system [5] is built upon the elemental EAM potentials for Al [6] and Cu [7], both of which were developed using large fitting databases combining experimental data and ab initio LAPW structural energies across a wide range of crystalline configurations. For the Al potential, the lattice parameter (4.05 Å vs. experimental 4.05 Å [2]), elastic constants ($c_{11} = 1.14 \times 10^{11}$ Pa vs. experimental $1.14 \times 10^{11}$ Pa [8], $c_{12} = 0.616 \times 10^{11}$ Pa vs. experimental $0.619 \times 10^{11}$ Pa [8], $c_{44} = 0.316 \times 10^{11}$ Pa vs. experimental $0.316 \times 10^{11}$ Pa [8]), intrinsic stacking fault energy (146 mJ/m² vs. experimental 166 mJ/m² [9]) are all in close agreement with experimental data. For the Cu potential, the lattice parameter (3.615 Å vs. experimental 3.615 Å [2]), elastic constants ($c_{11} = 1.699 \times 10^{11}$ Pa vs. experimental $1.700 \times 10^{11}$ Pa [8], $c_{12} = 1.226 \times 10^{11}$ Pa vs. experimental $1.225 \times 10^{11}$ Pa [8], $c_{44} = 0.762 \times 10^{11}$ Pa vs. experimental $0.758 \times 10^{11}$ Pa [8]), and intrinsic stacking fault energy (44.4 mJ/m² vs. experimental 45 mJ/m² [10]) are all in close agreement with experimental data.

**Supplementary Note 2: Effect of dissolution on LC strengthening**

LCs are only thermodynamically-stable in the presence of the dislocation and its stress field, meaning that they are in a metastable configuration after the first dislocation unpinning event. As such, there should be some tendency to return to a solid solution state that is favored by the bulk material and dissolution could potentially reduce the size of the particles over time. Our primary study simulates LCs that have similar sizes to those observed by Howard et al. [11]. These authors did not measure any significant difference in LC size when located at or away from the dislocations, suggesting that very little or no dissolution occurred. However, we still explore the possibility of LC dissolution here, where a controlled size reduction procedure was applied to the equilibrated LC configurations in both alloy systems (**Figure S1**). A cylindrical region centered on the LC and aligned along the dislocation line direction was defined and dopant atoms outside this cylinder were converted to bulk host atoms, effectively reducing the complexion volume while preserving the dislocation structure, simulation cell geometry, and all other parameters.

To enable a fair comparison between the two alloy systems, whose complexions have different forms, a complexion size reduction of ~45% was applied to each spatial dimension. Applied to a three-dimensional geometry (complete decoration of the dislocation line in Ni-Al), this factor yields a volumetric reduction of $[1 - (45\%)^3] \approx 91\%$; applied to a two-dimensional geometry (platelet linear complexions in Al-Cu), it yields an real reduction of $[1 - (45\%)^2] \approx 80\%$. The exact resulting complexion sizes obtained after the controlled reduction procedure were 91.1% and 79.6% of their respective original values, in close agreement with these geometric targets. This approach ensures that the two systems experience a visually and geometrically comparable degree of size reduction, despite the difference in dimensionality of their complexion structures.

Another set of shear deformation simulations were then performed on the reduced-size configurations at slip plane distances of d = 0 Å and d = -6 Å for the favored dislocation orientation, following the same protocol described in the manuscript. As shown in **Tables S1** and **S2**, even after reducing the complexion volumes significantly, the critical shear stress remains well above the solid solution baseline in both systems at both tested distances. In the Ni-Al system, the reduced-size critical stress values of 0.700 GPa and 0.725 GPa at d = 0 Å and d = -6 Å, respectively, remain substantially higher than the solid solution limit of 0.4 GPa in **Table S1**. Similarly, in the Al-Cu system, the reduced-size values of 0.20 GPa and 0.06 GPa at d = 0 Å and d = -6 Å significantly exceed the solid solution limit (0.02 GPa) in **Table S2**. These results confirm that a significant strengthening effect would persist in both alloy systems even if some linear complexion dissolution occurred, although some reduction in the strengthening effect is observed.

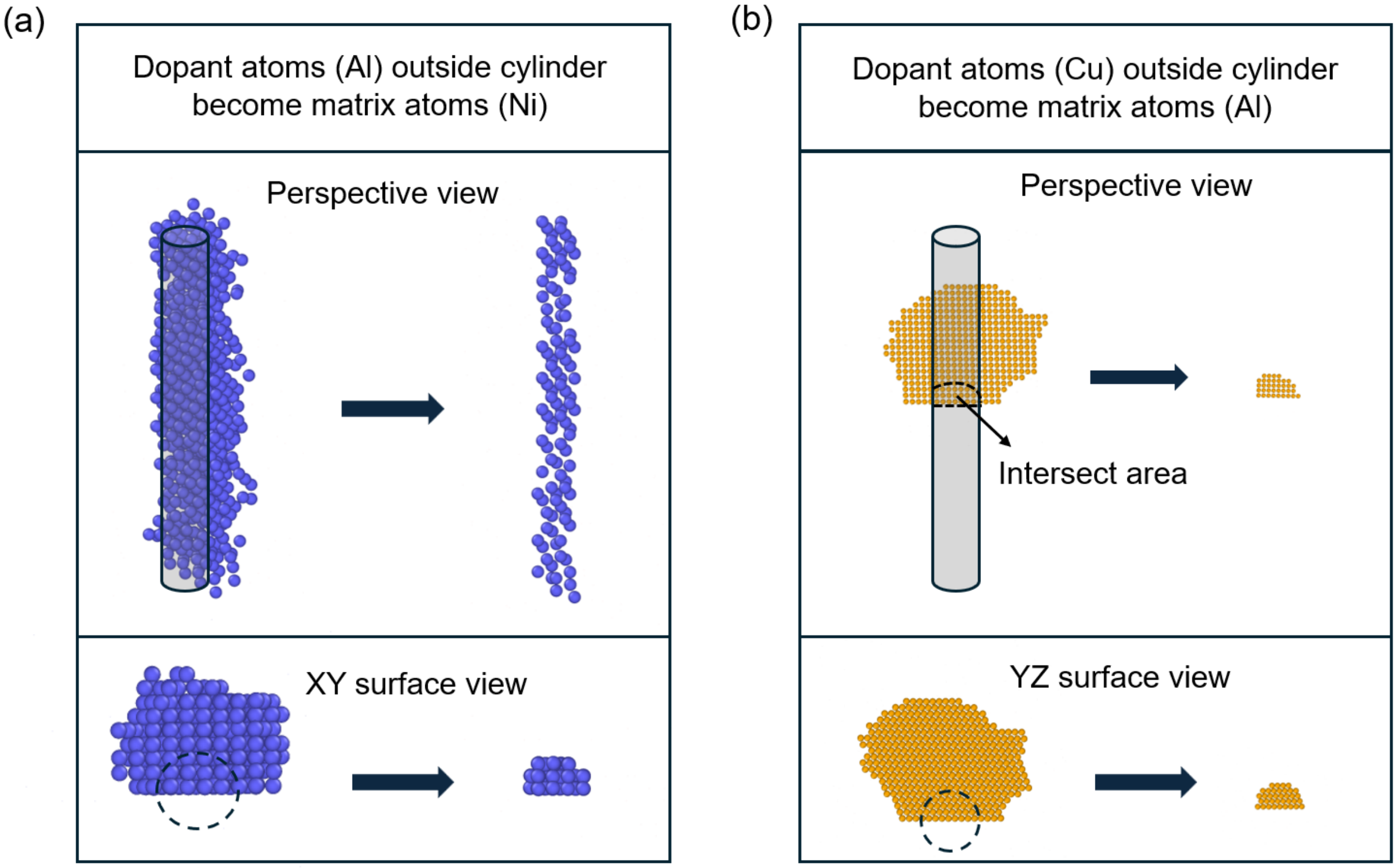

**Figure S1.** Schematic illustration of the complexion size reduction procedure applied to both alloy systems. (a) $L1_2$ nanoparticle array LC in Ni-Al before and after size reduction. (b) Platelet array LC in Al-Cu before and after size reduction.

**Table S1**. Critical shear stress of the Ni-Al system with and without LCs for original and reduced simulation sizes. The critical stress of the system without LC is 0.40 GPa.

| | Original size critical stress | Reduced size critical stress |
|---|---|---|
| $d$ = 0 Å | 1.20 GPa | 0.70 GPa |
| $d$ = -6 Å | 0.80 GPa | 0.73 GPa |

**Table S2**. Critical shear stress of the Al-Cu system with and without LCs for original and reduced simulation sizes. The critical stress of the system without LC is 0.02 GPa.

| | Original size critical stress | Reduced size critical stress |
|---|---|---|
| $d$ = 0 Å | 0.32 GPa | 0.20 GPa |
| $d$ = -6 Å | 0.08 GPa | 0.06 GPa |

## References:

[1] Y. Mishin, M. Mehl, D. Papaconstantopoulos, Embedded-atom potential for B 2− NiAl, Physical review B 65(22) (2002) 224114.

[2] C. Kittel, P. McEuen, Introduction to solid state physics, John Wiley & Sons2018.

[3] F. Kayser, C. Stassis, The elastic constants of Ni3Al at 0 and 23.5 C, Physica status solidi (a) 64(1) (1981) 335-342.

[4] P. Veyssiere, R. Noebe, Weak-beam study of< 111> superlattice dislocations in NiAl, Philosophical Magazine A 65(1) (1992) 1-13.

[5] F. Apostol, Y. Mishin, Interatomic potential for the Al-Cu system, Physical Review B—Condensed Matter and Materials Physics 83(5) (2011) 054116.

[6] Y. Mishin, D. Farkas, M. Mehl, D. Papaconstantopoulos, Interatomic potentials for monoatomic metals from experimental data and ab initio calculations, Physical Review B 59(5) (1999) 3393.

[7] Y. Mishin, M.J. Mehl, D.A. Papaconstantopoulos, A.F. Voter, J.D. Kress, Structural stability and lattice defects in copper: Ab initio, tight-binding, and embedded-atom calculations, Physical Review B 63(22) (2001) 224106.

[8] G. Simmons, Single crystal elastic constants and caluculated aggregate properties, A handbook 4 (1971).

[9] L.E. Murr, Interfacial phenomena in metals and alloys, (1974).

[10] C. Carter, I. Ray, On the stacking-fault energies of copper alloys, Philosophical Magazine 35(1) (1977) 189-200.

[11] H. Howard, W. Cunningham, A. Genc, B. Rhodes, B. Merle, T. Rupert, D. Gianola, Chemically ordered dislocation defect phases as a new strengthening pathway in Ni–Al alloys, Acta Materialia 289 (2025) 120887.